\documentclass[nocopyright]{cimento}

\usepackage{graphicx}  
\usepackage{lineno}

\title{Dielectrons at the LHC -- Chances and challenges}
\author{Harald~Appelsh\"auser (ALICE Collaboration)}
\instlist{\inst Goethe-Universit\"at Frankfurt - Frankfurt am Main, Germany}

\begin{document}

\newcommand{\mee}{m_{\rm ee}}
\newcommand{\ptee}{p_{\rm T,ee}}
\newcommand{\roots}{\sqrt{s}}
\newcommand{\rootsnn}{\sqrt{s_{\rm NN}}}
\newcommand{\ccbar}{c\bar{c}}
\newcommand{\bbbar}{b\bar{b}}
\newcommand{\Rppb}{R_{\rm pPb}}
\newcommand{\pt}{p_{\rm T}}
\newcommand{\dcaee}{{\rm DCA}_{\rm ee}}
\newcommand{\dcaeexyz}{{\rm DCA}_{\rm ee}^{\rm 3D}}

\maketitle

\begin{abstract}
Dielectrons are unique observables in ultra-relativistic heavy-ion collisions. Thanks to their penetrating nature, they carry information from all stages of the collision and can provide knowledge about pre-equilibirium dynamics, QGP temperature and transport coefficients, and chiral symmetry restoration. On the other hand, experimental challenges are enormous because production cross sections are small and the signal of interest is eclipsed by a huge combinatorial and physics background from light- and heavy-flavour hadron decays. In this talk the status of dielectron measurements with ALICE is shown and the perspectives with the recently installed and planned ALICE detector upgrades are discussed.
 
\end{abstract}

\section{Introduction}
Owing to their electromagnetic nature and correspondingly large mean free path in strongly interacting matter, photons and dielectrons are unique tools in the study of the properties of hot and dense
QCD matter created in high-energy heavy-ion collisions~\cite{ALICE_wp}, in particular the quark-gluon plasma (QGP). 
They are generated at all stages of the system evolution~\cite{gale1} and transport  
undisturbed information from the early, hot phase of the reaction into the final state. 
Beyond thermal radiation from quark-gluon plasma and the subsequent dense hadronic system~\cite{rapp1,rapp2}, photons and dielectrons are also considered as important
messengers of the very early stage of the collision, characterizing the dynamics towards equilibration~\cite{coq1,coq2}. 

On the other hand, a conceptual complication arises from the fact that the observed electromagnetic radiation is integrated over the different collision stages and needs to be disentangled in the final state. In this respect, there is a clear advantage for dielectrons since, in contrast to photons, they carry mass and therefore additional kinematic information that allows the contributions of different phases to be separated~\cite{rapp1}. While radiation  from the QGP  dominates the invariant mass ($\mee$) range above 
1~GeV/$c^2$, contributions below emerge predominantly from the hadronic phase. This allows  the properties of the QGP to be examined, these properties include its early temperature, equation of state and effective degrees of freedom, as well as the properties of hadrons in the medium, which currently represents the most promising experimental approach to exploring the chiral properties of the QCD phase transition~\cite{pisarski,rapp3}.

On the technical side, dielectrons suffer from lower production rates than photons due to an extra factor of the electromagnetic coupling constant $\alpha_{\rm EM}$. On the other hand, photon measurements are  limited by systematic uncertainties arising from the huge  background from electromagnetic decays of short-lived hadrons. Hadronic backgrounds to dielectrons are large as well, in particular at the highest collider energies, but arise mainly from correlated weak decays of hadrons with charm or beauty quarks. Experimental means to suppress these contributions are being developed.  

The thermal dielectron yield is determined by the space-time integral over the thermal emission rate described by the McLerran-Toimela formula~\cite{mcltoi}:
\begin{equation}
\frac{{\rm d}N_{\rm ee}}{{\rm d}^4x{\rm d}^4q}=-\frac{\alpha^2}{\pi^3 \mee^2}f^{BE}(q_0,T){\rm Im}\Pi_{EM}(\mee,q,\mu_B,T).
\end{equation}
For a structureless spectral function ${\rm Im}\Pi_{EM}$, as is the case at dielectron masses above the hadron-parton duality threshold at $\mee \approx 1$~GeV/$c^2$, the dielectron yield follows a simple exponential given by the 
Bose-Einstein weight $f^{BE}(q_0,T)$. In this mass region, the dielectron yield is dominated by radiation from the early phase due to  its higher temperature and allows a straight-forward determination of the effective QGP temperature by an exponential fit to the data. At lower masses, radiation from the hadronic phase dominates due to the much larger space-time volume and the spectral function exhibits a non-trivial shape, depending on $T$ and $\mu_B$. Below $T_c$ its properties reflect hadronic degrees of freedom in finite $T$ and $\mu_B$ that are related to chiral symmetry restoration. Extraction of the medium temperature from low-mass data therefore requires additional assumptions, i.e. that the spectral function at $\mee < 1$~GeV/$c^2$ becomes structureless as well in the relevant range of $T$ and $\mu_B$.

Heavy-ion collisions at the LHC produce the hottest, largest, and longest-lived system to date which makes the LHC the ideal place to investigate the properties of the QGP. On the other hand, dielectron studies at the LHC are complicated by huge combinatorial and physical backgrounds. In particular, the large cross section for charm and beauty production and the large branching ratios into correlated semi-leptonic weak decays pose a huge challenge for analysis. In the so-called intermediate-mass region (IMR) between the $\phi$- and $J/\psi$-meson peaks, the contributions from heavy-flavor decays outshine the expected thermal dielectron yield by about one order of magnitude. A very precise knowledge of the heavy-flavour contributions is therefore essential for any thermal dielectron analysis at collider energies. In this presentation, the current experimental situation at the LHC is reviewed and the prospects for future dielectron measurements with ALICE are discussed.

\section{Charm and beauty production - constraints from pp and p--Pb collisions}

\begin{figure}
\includegraphics[scale=0.33]{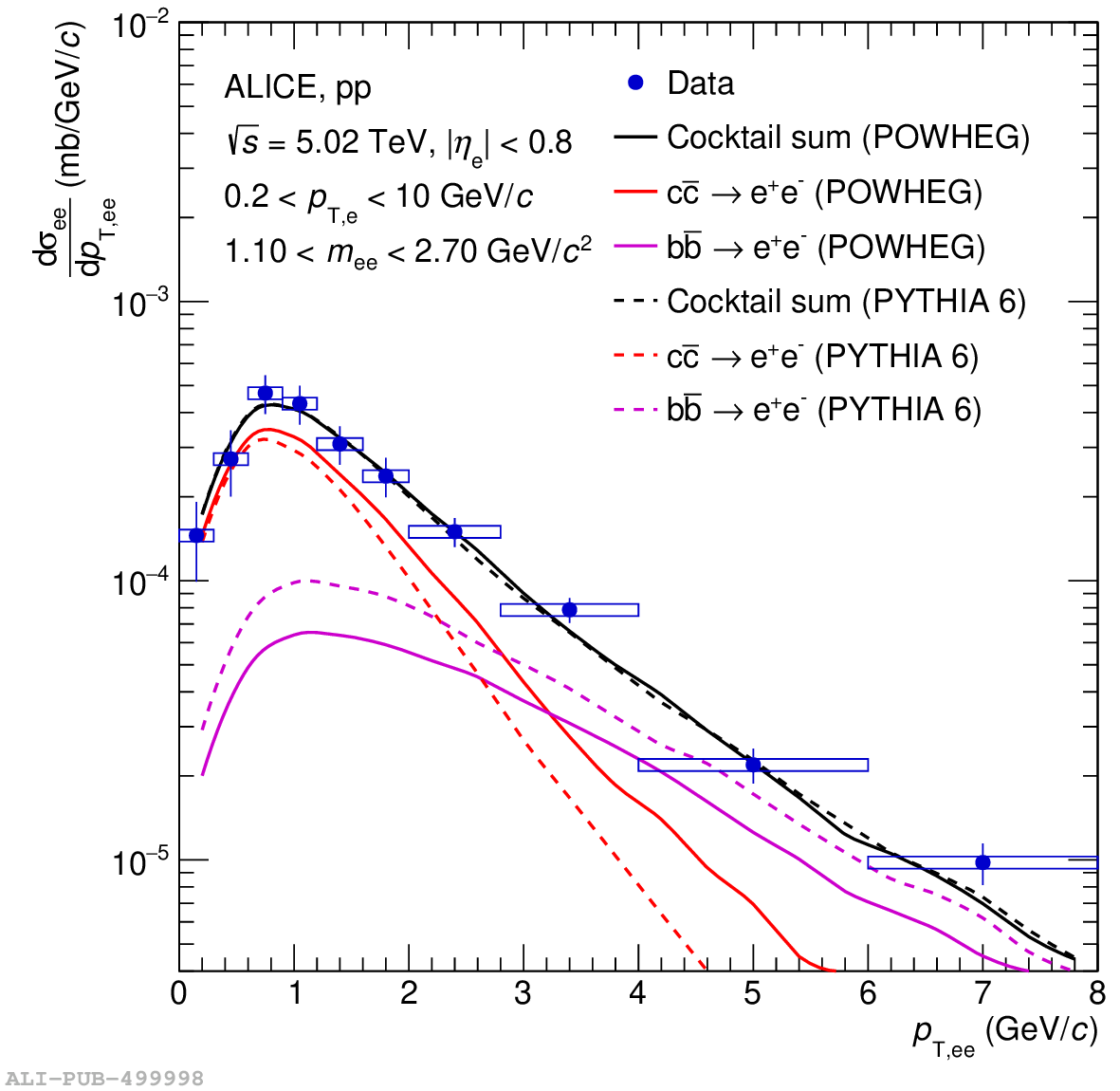}     
\includegraphics[scale=0.33]{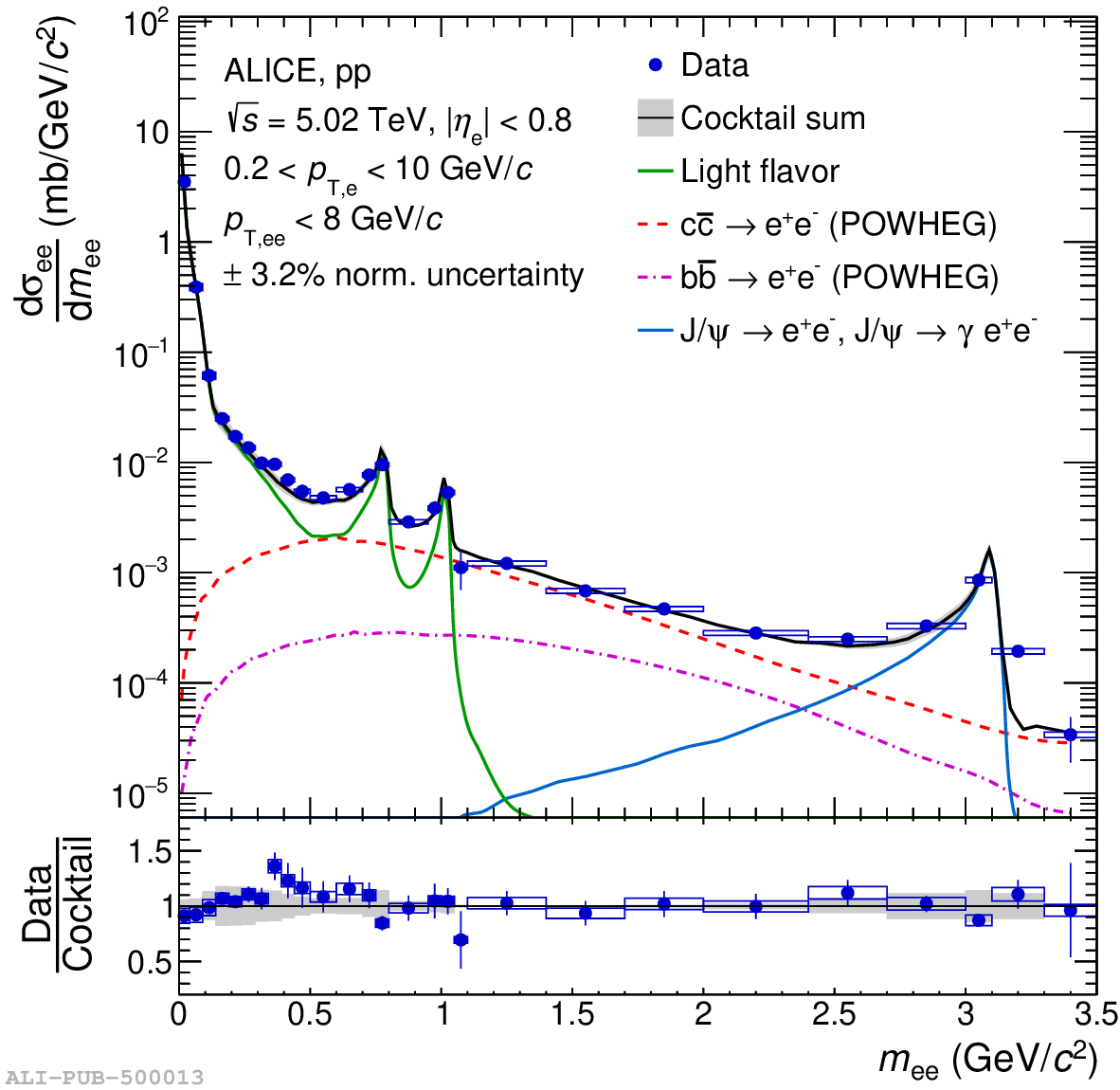}
\caption{Dielectron $\ptee$ (left) and $\mee$ (right) distributions in pp collisions at $\roots=5.02$~TeV together with the fitted contributions from $\ccbar$ and $\bbbar$ decays (from~\cite{alice-pp-pPb}). }
\label{fig1}
\end{figure}

Constraints on charm and beauty production in lead-lead (Pb--Pb) collisions can be derived from measurements in smaller collision systems. Dielectron reference data from proton-proton (pp) and proton-lead (p--Pb) collisions at the same centre-of-mass energy $\rootsnn$ and with the same experimental setup are particularly useful because sytematic uncertainties arising from uncertainties of branching ratios and phase-space extrapolations cancel. ALICE~\cite{alice-jinst,alice-run12} has conducted a comprehensive dielectron program in pp, p--Pb, and Pb--Pb collisions at $\rootsnn=5.02$~TeV from LHC 
Run~2~\cite{alice-pp-pPb,alice-PbPb} .

Dielectron yields in pp collisions at 
$\rootsnn=5.02$~TeV are shown in Fig.\,\ref{fig1}. The left panel shows the dielectron-pair transverse-momentum ($\ptee$) distribution in the IMR where, in pp collisions, heavy-flavour decays are expected to dominate the yield by far. A combined fit of $\ccbar$ and $\bbbar$ distributions, modeled by PYTHIA6~\cite{pythia6} and POWHEG~\cite{pow1,pow2,pow3,pow4}, yield slightly different relative contributions from charm and beauty for the two generators, but both describe the inclusive $\ptee$ spectrum equally well.
The right panel shows the same data as a function of $\mee$ compared to POWHEG where also contributions from light-flavour mesons (green) and $J/\psi$ (blue) are included~\cite{alice-pp-pPb}. A very good description of the measured dielectron yield is observed. The $\ccbar$ and $\bbbar$ yields obtained from the fit of POWHEG templates to pp data are used as input for the construction of the so-called hadronic cocktail in p--Pb and Pb--Pb collisions, as shown in the following.

\begin{figure}
\includegraphics[scale=0.33]{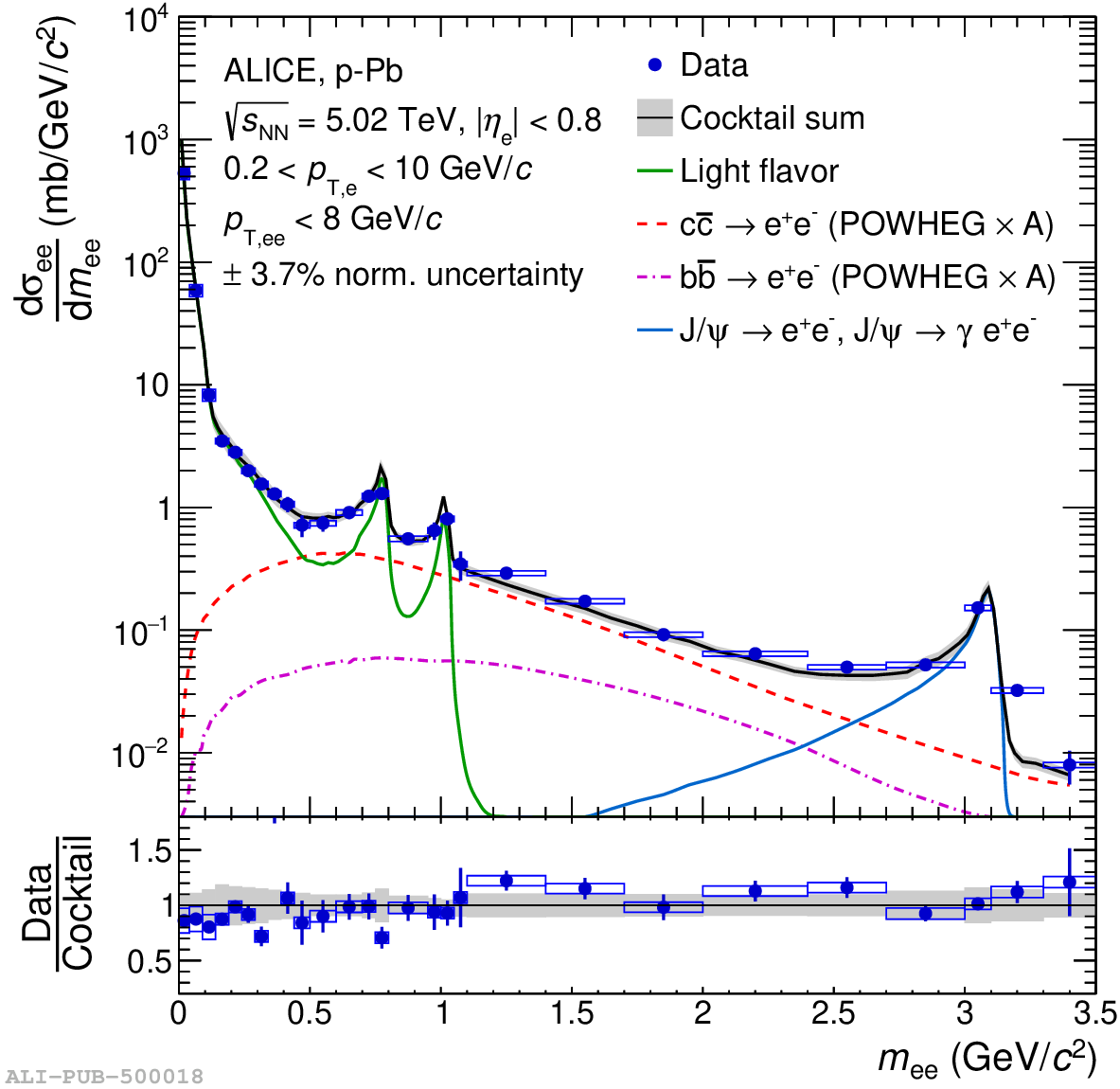}     
\includegraphics[scale=0.33]{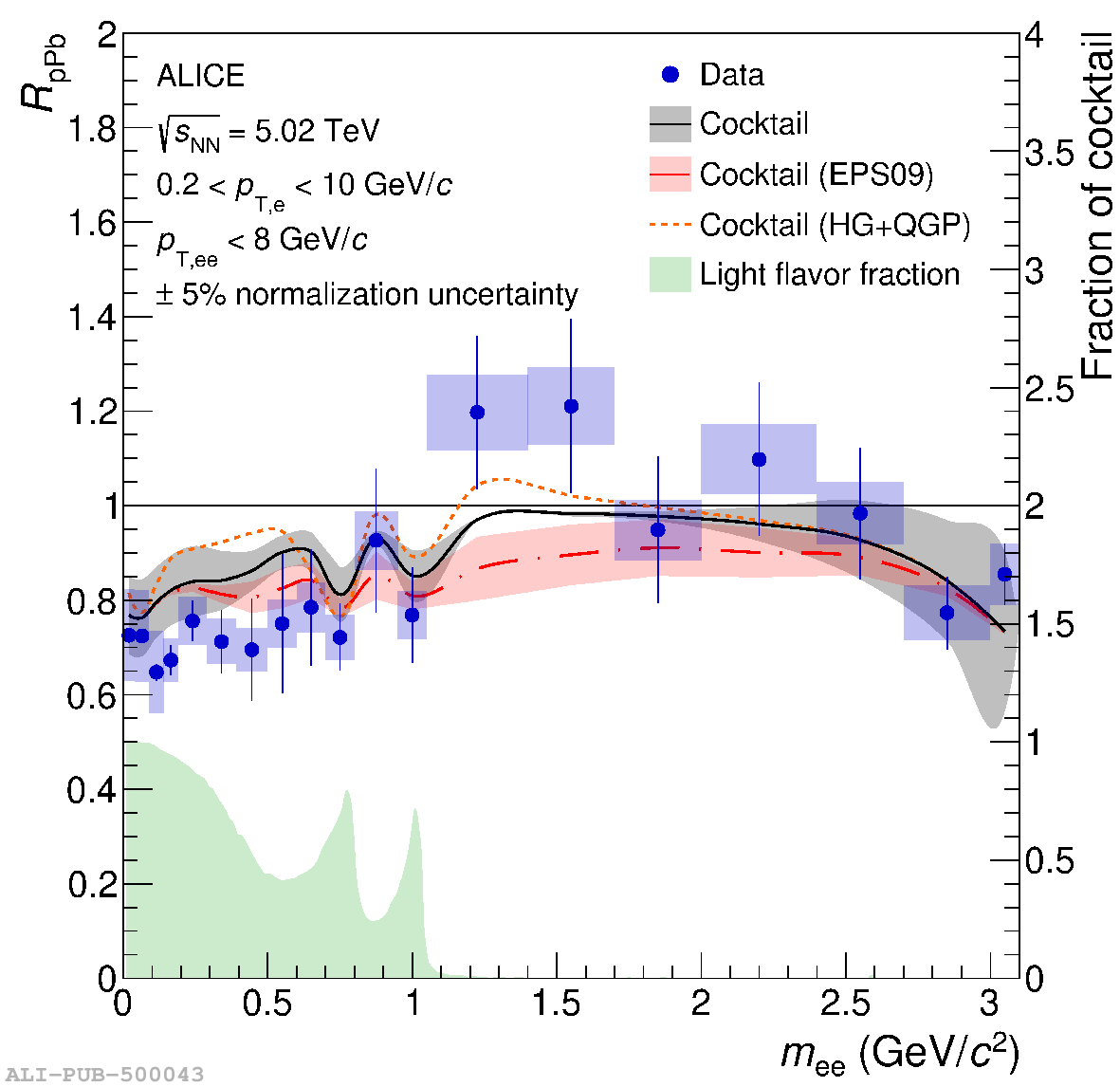}
\caption{Left: dielectron $\mee$ distributions in p--Pb collisions at $\rootsnn=5.02$~TeV
compared to a cocktail of expected hadronic decay contributions. Right: dielectron $\Rppb$
compared to different cocktails and a calculation including a thermal dielectron 
contribution (see text)(from~\cite{alice-pp-pPb}).}
\label{fig2}
\end{figure}

Figure~\ref{fig2} (left panel) shows the dielectron invariant mass 
spectrum in p--Pb collisions at $\rootsnn = 5.02$~TeV~\cite{alice-pp-pPb}. The data are compared to a hadronic decay cocktail composed of contributions 
from light-flavour hadrons (green), $J/\psi$ (blue) and the POWHEG-based fitted contribution measured in pp (see above), scaled by the mass number $A=208$. The description is reasonable within systematic uncertainties, however, a slight tension is observed at low and intermediate mass.

The nuclear modification factor $\Rppb$
\begin{equation}
\Rppb(\mee) = \frac{1}{A}\frac{{\rm d}\sigma_{\rm ee}^{\rm pPb}/{\rm d}\mee}
{{\rm d}\sigma_{\rm ee}^{\rm pp}/{\rm d}\mee}
\end{equation}
is shown on the right panel of Fig.\,\ref{fig2} together with the corresponding ratio of the cocktails~\cite{alice-pp-pPb}. 
A suppression below unity is expected at low mass because the dominant light-flavor contribution in the cocktail does not scale with $A$. Below 
$\mee \approx 1$~GeV/$c^2$, however, the data tend to be below the cocktail, possibly indicating cold nuclear matter (CNM) 
effects in the Pb-nucleus. 
A cocktail based on nuclear parton distribution functions from EPS09~\cite{eps09} yields a better description of the data at low mass, however, it slightly underpredicts the measured $\Rppb$ in the IMR. This may be compatible with a small additional contribution from thermal radiation, as shown by the dashed red line~\cite{rapp2}. The present dielectron data in p--Pb are compatible with a suppression observed in the $\Rppb$ of Dmesons below $\pt = 2$~GeV/$c$~\cite{aliceD1,aliceD2}. The clarification of a possible thermal dielectron contribution in 
p--Pb collisions requires higher-precision data that will be collected by the 
upgraded ALICE apparatus in LHC Run 3 and 4 (see below).

\section{Search for QGP radiation in Pb--Pb}
\begin{figure}
\includegraphics[scale=0.33]{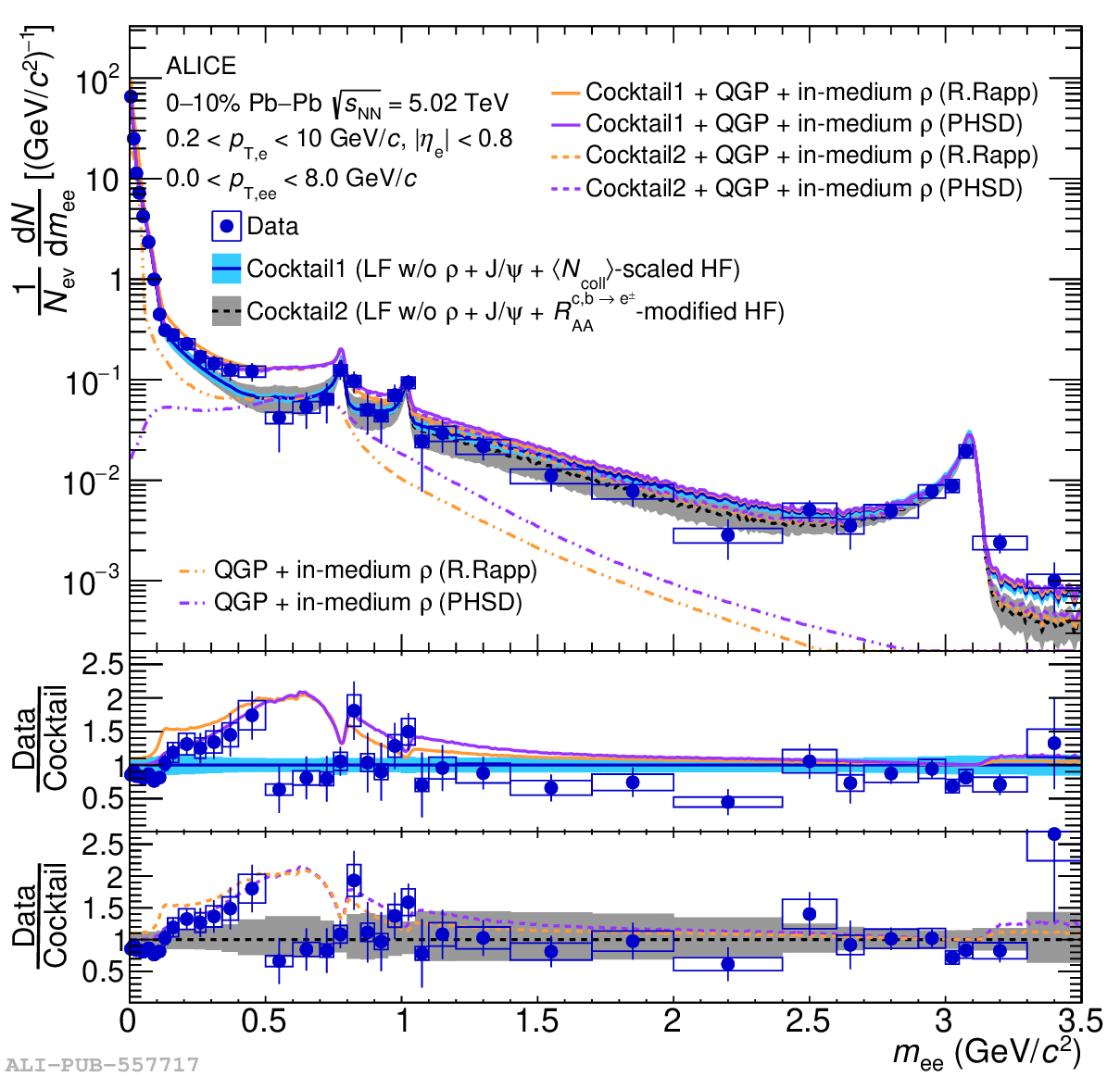}     
\includegraphics[scale=0.33]{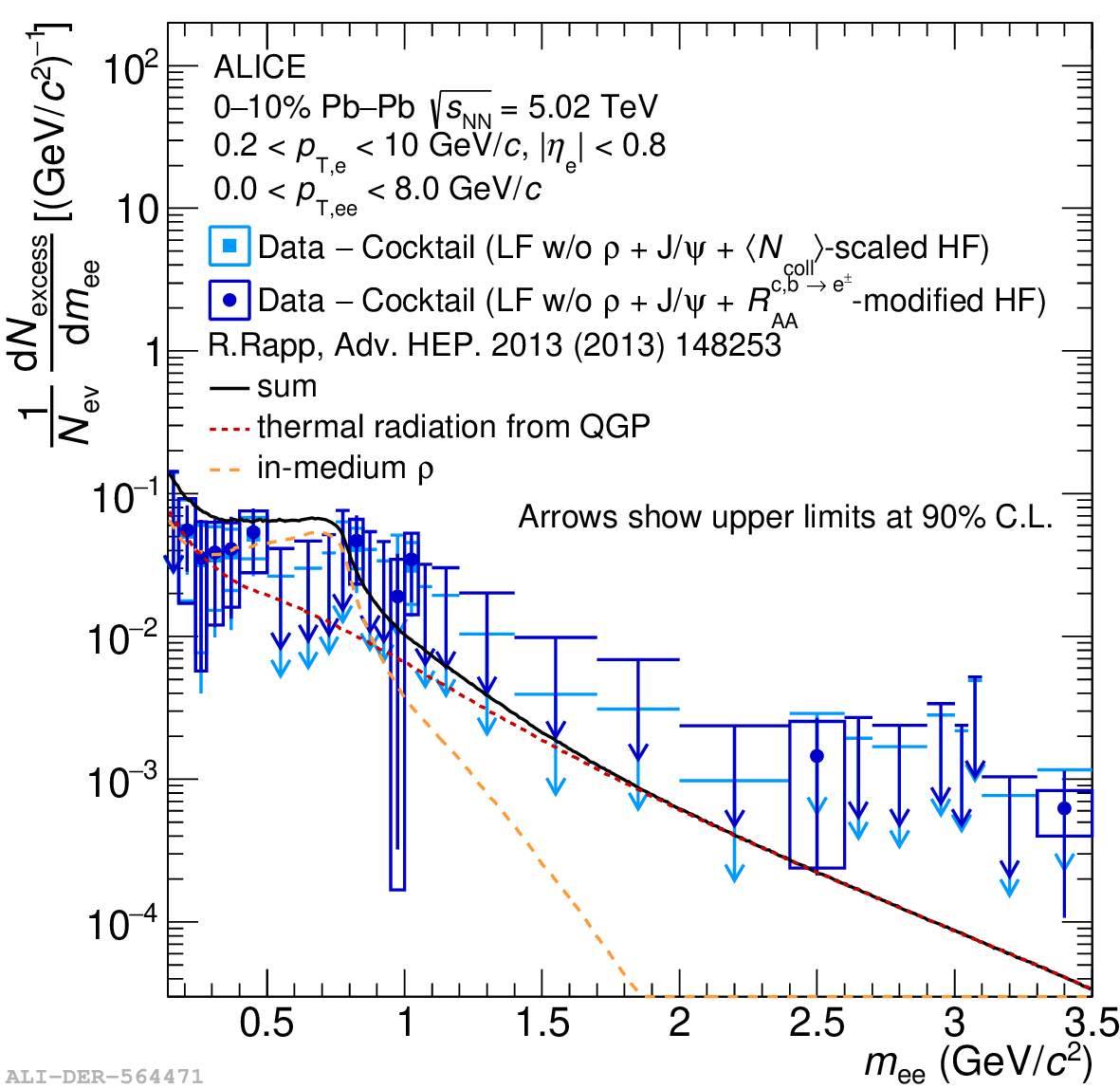}
\caption{Upper left: dielectron $\mee$ distributions in the 10\% most central Pb--Pb collisions at $\rootsnn = 5.02$~TeV, compared to two different hadronic decay cocktails and models (see text). 
Middle and bottom left: ratios of data and models to cocktails. Right: excess yield of dielectrons with respect to the
known hadronic sources compared to model calculations (from~\cite{alice-PbPb}).}
\label{fig3}
\end{figure}

Figure~\ref{fig3} shows recently published results of dielectron production in Pb--Pb collisions from LHC Run2 in 2018~\cite{alice-PbPb}. The upper left panel shows the $\mee$ spectrum in central (0-10\%) collisions together with a comparison to the expected yields from hadronic decay cocktails and model calculations. Cocktail~1 employs the measured heavy-flavour contribution in pp collision at the same $\rootsnn$ (see Fig.\,\ref{fig1}), scaled by the mass number $A=208$. In the middle panel, the ratio of data to cocktail is shown. In the low-mass region ($0.18<\mee < 0.5$~GeV/$c^2$) the data exceed the cocktail by a factor 
$1.42 \pm 0.12 (\rm{stat.}) \pm 0.17 (\rm{syst.}) \pm 0.12 (\rm{cocktail})$. The data are compatible with calculations from the PHSD transport model~\cite{phsd} and a hadronic many-body 
approach~\cite{rapp2} which 
indicates a strong in-medium modification of the 
$\rho$ spectral function at the LHC. In the mass range $0.5<\mee<1$~GeV/$c^2$ the data are compatible with the cocktail within large experimental uncertainties. In the IMR, however, the data tend to be below the cocktail.  
This finding may be compatible with CNM effects and final-state modifications of charmed hadrons in nuclear collisions, as observed in D-meson nuclear modification factors $R_{\rm AA}$~\cite{alice-D-pbpb-raa}. In Cocktail~2, nuclear effects are taken into account by folding Cocktail~1 with initial- and final-state effects from EPS09 and measured heavy-flavour electron $R_{\rm AA}$ results from central Pb--Pb collisons~\cite{alice-hfe-raa}, respectively. Comparison to data yields a better agreement than in the case of Cocktail~1 (see Fig.\,\ref{fig3} lower left panel). However,
the procedure to include nuclear effects from EPS09 and Pb--Pb data gives rise to a sizeable increase of the systematic uncertainties of the cocktail, rendering the observation of possible excess due to thermal dielectron radiation from the QGP impossible. This is demonstrated in the right panel of 
Fig.\,\ref{fig3} where Cocktail~2 is subtracted from the data. 
Cocktail-agnostic strategies must therefore be persued to isolate thermal QGP radiation at the LHC.

\section{Topological separation of dielectron sources}
\begin{figure}
\includegraphics[scale=0.33]{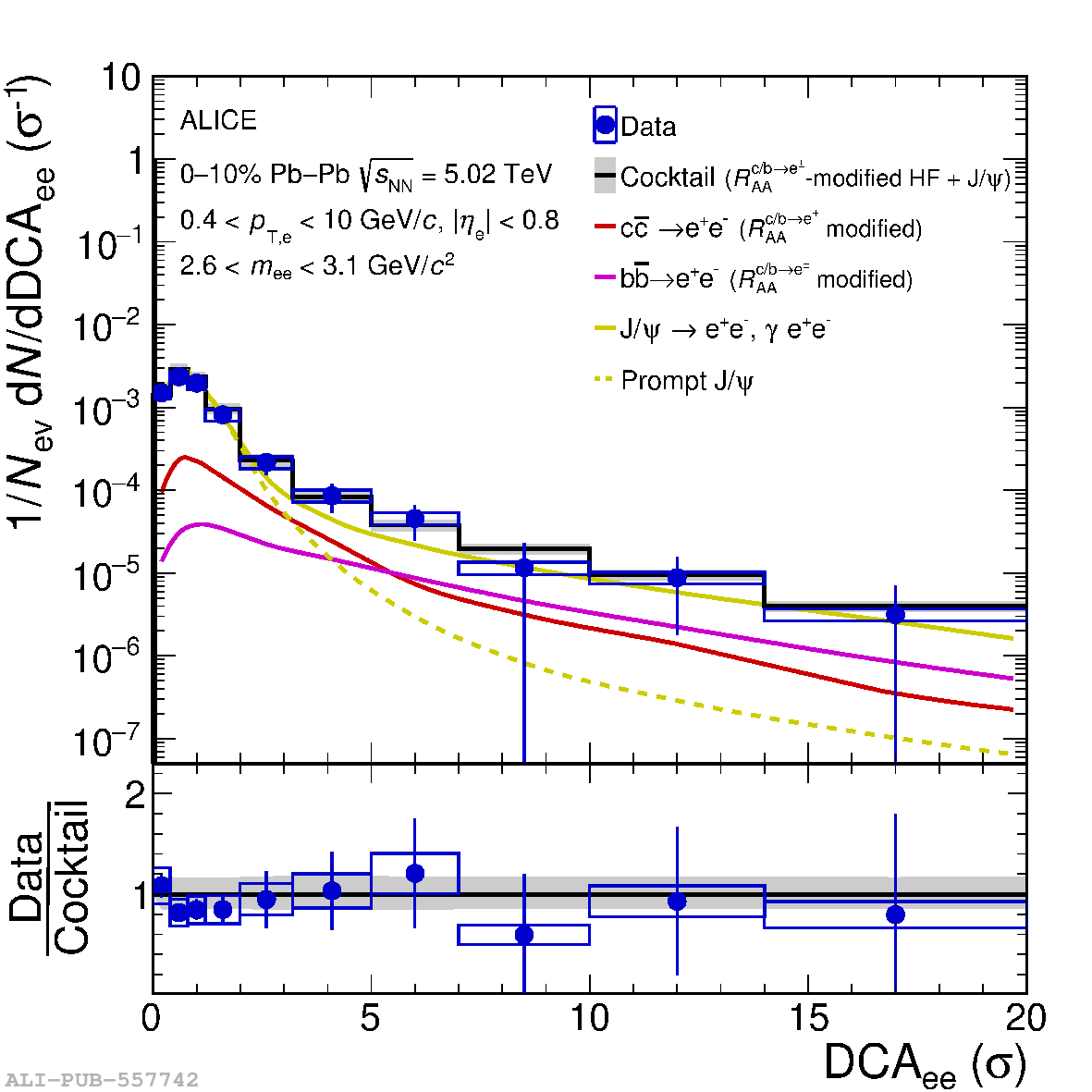}     
\includegraphics[scale=0.33]{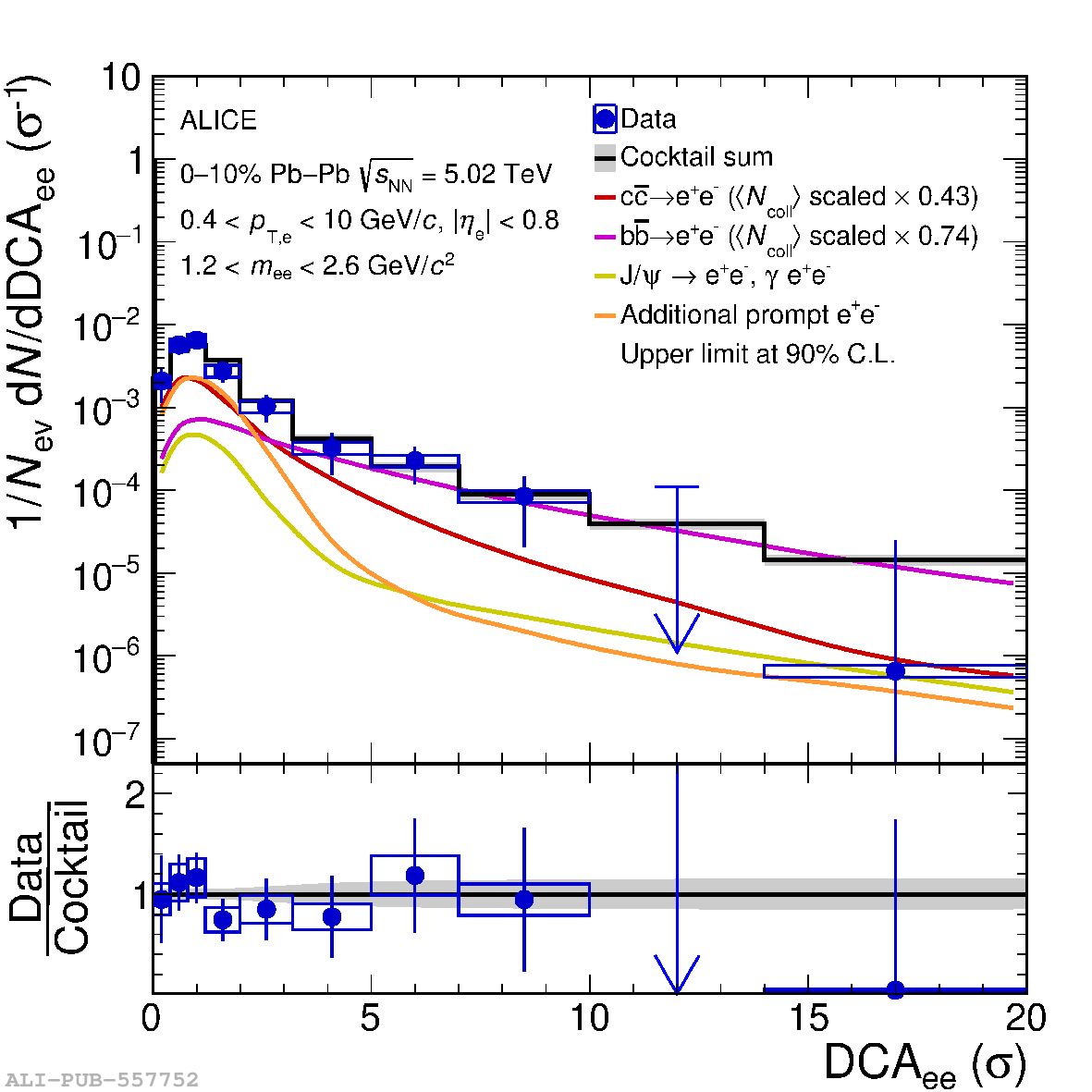}
\caption{Left: dielectron yield in the 10\% most central Pb--Pb collisions at $\rootsnn = 5.02$~TeV as a function of $\dcaee$ in $2.6 < \mee < 3.1$~GeV/$c^2$, compared with a cocktail of expected sources. Right: fit of the dielectron yield in $1.2 < \mee < 2.6$~GeV/$c^2$. Bottom panels: ratios of data to cocktail.}
\label{fig4}
\end{figure}

Typical weak decay lengths of a few
hundred $\mu$m of hadrons with charm or beauty quarks can be measured 
with Si-based vertex detectors such as the Inner Tracking System (ITS)~\cite{alice-jinst} 
of ALICE. To this end, the normalized pair distance-of-closest-approach ($\dcaee$) is evaluated for each dielectron candidate:
\begin{equation}
\dcaee = \sqrt{\frac{{\rm DCA}_1^2+{\rm DCA}_2^2}{2}}
\end{equation}
where ${\rm DCA}_1$ and ${\rm DCA}_2$ are the distances-of-closest-approach of
the electron and positron tracks to the main collision vertex in the transverse plane. This technique was pioneered by the NA60 collaboration for dimuons in the forward hemisphere~\cite{na60-dca}. Recently, a first $\dcaee$ analysis 
with dielectrons at midrapidity was
performed at the LHC, see Fig.\,\ref{fig4}~\cite{alice-PbPb}. The left panel
shows the distribution $\dcaee$, normalized to the resolution $\sigma$, in the mass region around the $J/\psi$. 
Also shown are the expectations from the hadronic cocktail, where the 
shapes of the $\dcaee$ distributions are generated by a Monte-Carlo (MC) simulation. The cocktail contains prompt and non-prompt $J/\psi$, and correlated decays from $\ccbar$ and $\bbbar$ pairs including initial- and final-state 
modifications as implemented in Cocktail~2. Data and cocktail agree well over the full measured $\dcaee$ range. In the right panel of Fig.\,\ref{fig4} the $\dcaee$ distribution in the IMR is shown. While the $J/\psi$ yield is
constrained from data, the yields from $\ccbar$ and $\bbbar$ are now unconstrained 
from cocktails. In addition, a prompt dielectron contribution is included in the fit.
The best fit implies a reduction by a factor $0.43 \pm 0.4 (\rm{stat.}) \pm 0.22 (\rm{syst.})$ for $\ccbar$ and $0.74 \pm 0.24 (\rm{stat.}) \pm 0.12 (\rm{syst.})$ for $\bbbar$ with respect
to the expectation from Cocktail~1. At the same time, the fit favours a finite prompt 
contribution that yields $3.17 \pm 3.81 (\rm{stat.}) \pm 0.35 (\rm{syst.})$ 
and $1.15 \pm 1.38 (\rm{stat.}) \pm 0.13 (\rm{syst.})$ times
the contribution from the hadronic many-body and PHSD calculations, respectively. For
the first time at the LHC a hint for a prompt thermal dielectron component is observed in the IMR with small systematic uncertainties, albeit with a statistical significance of only $1\sigma$. Nevertheless, this study opens the door for the next generation
of high-precision measurements with the upgraded ALICE detector in Runs~3 and 4. 

\section{First dielectron measurements with ALICE in Run~3}
In preparation of the LHC Runs 3 and 4 
the ALICE apparatus underwent a 
significant upgrade. 
The TPC, which is main tracking and electron identification detector, was equipped with a GEM-based, continuous readout system~\cite{tpc-gem}, allowing an increase of the readout rate by a factor 100 (in Pb--Pb) to 
1000 (in pp) with respect to Run~1 and 2. 
A new all-pixel inner tracking system (ITS2)
was installed which improves the pointing resolution of 
charged-particle tracks by a factor 3 (5) in the $xy$- ($z$-) direction~\cite{ITS2}. This enables a 3D definition of the dielectron DCA, $\dcaeexyz$, 
which enhances significantly the capabilities to seperate 
prompt dielectrons from heavy-flavour decays.

\begin{figure}
\includegraphics[scale=0.33]{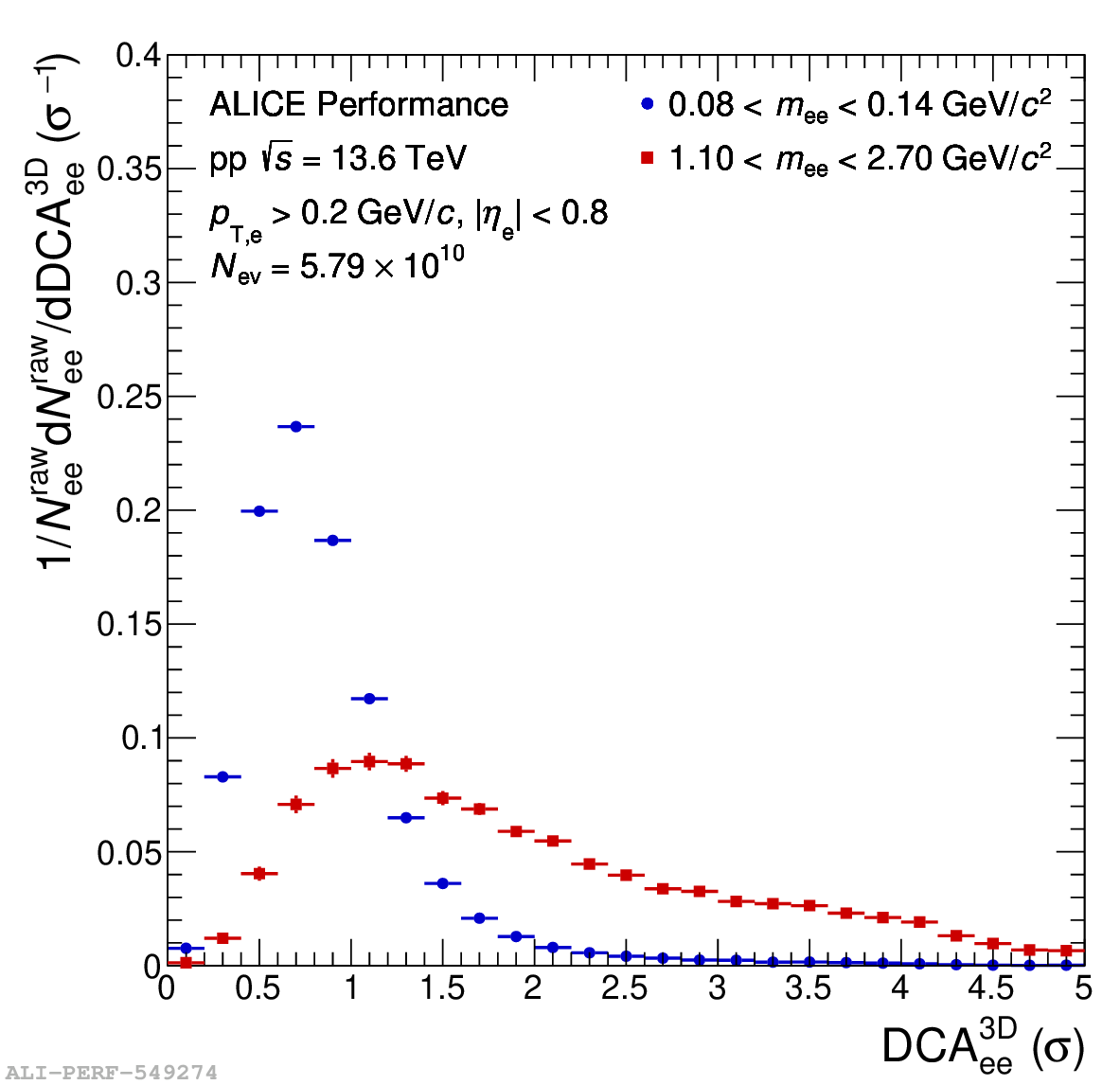}     
\includegraphics[scale=0.33]{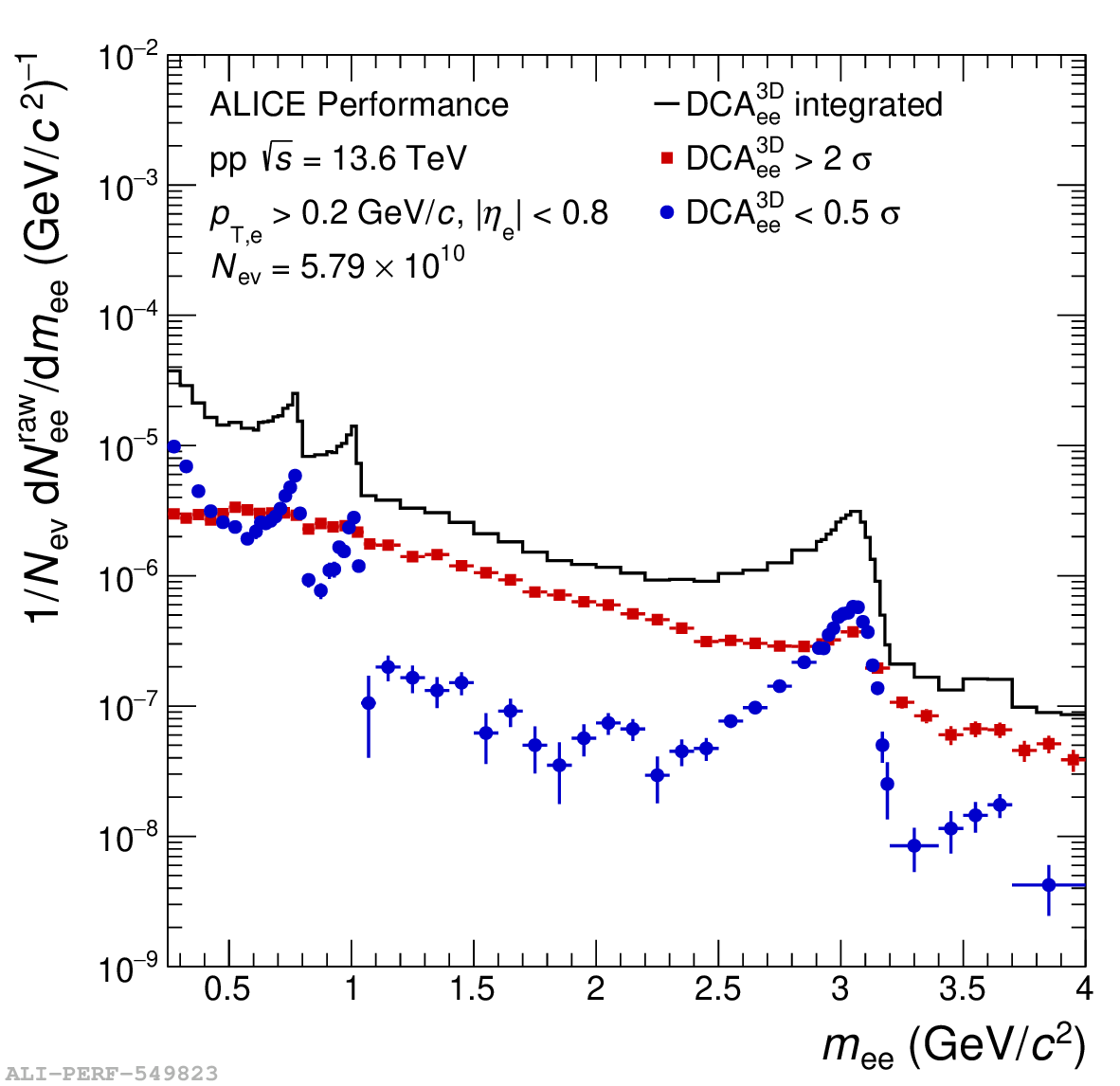}
\caption{Left: $\dcaeexyz$ distributions in different $\mee$ intervals in pp 
collisions at $\roots = 13.6$~TeV from Run~3. Right: inclusive raw dielectron yield (black histogram) and after cuts in $\dcaeexyz$ (blue and red) as a 
function of $\mee$.}
\label{fig5}
\end{figure}

First dielectron measurements made with 1~pb$^{-1}$ of pp collision data, recorded with the upgraded ALICE detector in Run~3, are shown in Fig.\,\ref{fig5}. The left panel shows $\dcaeexyz$ in two different
$\mee$ intervals. The blue distribution ($0.08 < \mee < 0.14$~GeV/$c^2$) is dominated by 
prompt Dalitz decays of $\pi^0$-mesons, while the red distribution ($1.1 < \mee < 2.7$~GeV/$c^2$)
is populated almost exclusively by heavy-flavour decays. While the blue distribution represents the detector resolution, the red distribution exhibits a clear tail owing to 
the finite lifetime of the hadrons which can be resolved with the ITS2. The effect of tight
cuts on $\dcaeexyz$ on the inclusive $\mee$ spectrum is shown in the right panel of Fig.\,\ref{fig5}. While a selection on $\dcaeexyz < 0.5\sigma$ enhances the prompt contributions from Dalitz decays and vector mesons, including the $\psi'$ (blue points), a selection 
$\dcaeexyz > 2\sigma$ produces the expected continuum from heavy-flavour decays and a peak 
from non-prompt $J/\psi$ from $b$-decays. This demonstrates the unprecedented precision
in the pointing resolution of ITS2 and highlights the large potential for isolating the 
prompt thermal dielectron contribution in the IMR from the huge background from hadron decays.

\section{Outlook}
The upgraded ALICE detector at the LHC will enable high-precision  dielectron measurements in pp, p--Pb, and Pb--Pb collisions thanks to its improved pointing resolution and
increased readout rate via continuous readout. A further increase of the sensitivity
will be achieved by a new vertex tracker (ITS3) for Run~4 which improves the pointing resolution by another factor of about~3~\cite{its3}. These upgrades will allow for a precise discrimination of prompt and non-prompt dielectrons, enabling a comprehensive study of in-medium modifications
of hadrons and its connection to chiral symmetry restoration, and an exploration 
of the early temperature of the system. Furthermore, a measurement of azimuthal 
anisotropies of thermal dielectrons can give access to early collectivity of
the system and the partonic equation of state. Polarization measurements at 
high invariant masses between $J/\psi$ and $\Upsilon$ may yield a better understanding
of early equilibration processes and provide an independent assessment of the 
specific viscosity $\eta/s$ of the partonic system~\cite{coq1,coq2}. Dedicated runs at reduced
magnetic field will also consolidate recent hints for a new dielectron 
source in pp collisions at very low $\mee$ and $\ptee$~\cite{alice-pp-lowb}.

A next-generation heavy-ion experiment ALICE~3 is proposed to continue precision studies of a large number of observables during LHC Runs 5 and 6. In the dielectron sector, unprecedented rate capability,  pointing resolution and sensitivity at very low $\pt$ will enable precision studies of the 
time evolution of the early temperature and collectivity, and spectral modifications 
related to chiral symmetry restoration~\cite{alice3-loi}. Furthermore, there will be access to yet unexplored phenomena 
at very low $\pt$ such as the emergence of soft dielectrons connected to the 
Low theorem~\cite{low} and a possible measurement of the electric conductivity of the QGP~\cite{cond}.







%



\begin{thebibliography}{0}
\bibitem{ALICE_wp} 
	\BY{ALICE Collaboration} 
	\TITLE{The ALICE experiment -- A journey through QCD}, 
	arXiv:2211.04384 [nucl-ex].
\bibitem{gale1} 
	\BY{Gale A.} 
	\TITLE{Photon Production in Hot and Dense Strongly Interacting Matter},
	\IN{Landolt-Bornstein}{23}{2010}{445}, 
	arXiv:0904.2184 [hep-ph].
\bibitem{rapp1} 
	\BY{Rapp R. \atque van Hees H.} 
	\TITLE{Thermal Dileptons as Fireball Thermometer and Chronometer}, 
	\IN{Phys. Lett. B}{753}{2016}{586-590}, 
	arXiv:1411.4612 [hep-ph].
\bibitem{rapp2} 
	\BY{Rapp R.} 
	\TITLE {Dilepton Spectroscopy of QCD Matter at Collider Energies}, 
	\IN{ Adv. High Energy Phys.}{2013}{2013}{148253},  
	arXiv:1304.2309 [hep-ph].
\bibitem{coq1} 
	\BY{Coquet M., Du X., Ollitraut J.-Y., Schlichting S., \atque Winn M.} 
	\TITLE{Intermediate mass dileptons as pre-equilibrium probes in heavy ion 		collisions}, 
	\IN{Phys. Lett. B}{821}{2021}{136626}, 
	arXiv:2104.07622 [nucl-th].
\bibitem{coq2} 
	\BY{Coquet M., Du X., Ollitraut J.-Y., Schlichting S., \atque Winn M.} 
	\TITLE{Transverse mass scaling of dilepton radiation off a quark-gluon plasma},
	\IN{Nucl. Phys. A}{1030}{2023}{122579}, 
	arXiv:2112.13876 [nucl-th].
\bibitem{pisarski}
	\BY{Pisarski R.D.}
	\TITLE{Where does the $\rho$ go? Chirally symmetric vector mesons in the quark-gluon plasma},
	\IN{Phys. Rev. D}{52}{1995}{R3773-R3776},
	arXiv:hep-ph/9503328.
\bibitem{rapp3}
	\BY{Hohler P.M. \atque Rapp R.}
	\TITLE{Is $\rho$-Meson Melting Compatible with Chiral Restoration?}
	\IN{Phys. Lett. B}{731}{2014}{103-109},
	arXiv:1311.2921 [hep-ph].
\bibitem{mcltoi}
	\BY{McLerran L.D. \atque Toimela T.}
	\TITLE{Photon and Dilepton Emission from the Quark - Gluon Plasma: Some General Considerations},
	\IN{Phys. Rev. D}{31}{1985}{545}
\bibitem{alice-jinst}
	\BY{ALICE Collaboration}
	\TITLE{The ALICE experiment at the CERN LHC},
	\IN{JINST}{3}{2008}{S08002}.
\bibitem{alice-run12}
	\BY{ALICE Collaboration}
	\TITLE{Performance of the ALICE Experiment at the CERN LHC},
	\IN{Int. J. Mod. Phys. A}{29}{2014}{1430044},
	arXiv:1402.4476 [nucl-ex].
\bibitem{alice-pp-pPb}
	\BY{ALICE Collaboration}
	\TITLE{Dielectron production in proton-proton and proton-lead collisions at $\rootsnn = 5.02$~TeV},
	\IN{Phys. Rev. C}{102}{2020}{055204}
\bibitem{alice-PbPb}	
	\BY{ALICE Collaboration}
	\TITLE{Dielectron production in central Pb--Pb collisions at $\rootsnn = 5.02$~TeV},
	arXiv:2308.16704 [nucl-ex].
\bibitem{pythia6}
	\BY{Sjostrand T, Mrenna S., \atque Skands P.Z.}
	\TITLE{PYTHIA 6.4 Physics and Manual},
	\IN{JHEP}{05}{2006}{026},
	arXiv:hep-ph/0603175.
\bibitem{pow1}
	\BY{Nason N.}
	\TITLE{A New method for combining NLO QCD with shower Monte Carlo algorithms},
	\IN{JHEP}{11}{2004}{040},
	arXiv:hep-ph/0409146 [hep-ph].
\bibitem{pow2}
	\BY{Frixione S., Nason P., \atque Ridolphi G.}
	\TITLE{A Positive-weight next-to-leading-order Monte Carlo for heavy flavour hadroproduction},
	\IN{JHEP}{09}{2007}{126}
\bibitem{pow3}
	\BY{Frixione S., Nason P., \atque Oleari C.}
	\TITLE{Matching NLO QCD computations with Parton Shower simulations: the POWHEG method},
	\IN{JHEP}{11}{2007}{070}
\bibitem{pow4}
	\BY{Frixione S., Nason P., Oleari C., \atque Re E.}
	\TITLE{A general framework for implementing NLO calculations in shower Monte Carlo programs: the POWHEG BOX},
	\IN{JHEP}{06}{2010}{043},
	arXiv:1002.2581 [hep-ph].
\bibitem{eps09}
	\BY{Eskola K.J., Paukkunen H., \atque Salgado C.A.}
	\TITLE{EPS09: A New Generation of NLO and LO Nuclear Parton Distribution Functions},
	\IN{JHEP}{04}{2009}{065},
	arXiv:0902.4154 [hep-ph].
\bibitem{aliceD1}
	\BY{ALICE Collaboration}
	\TITLE{D-meson production in p--Pb collisions at $\rootsnn =5.02$~TeV and in 	pp collisions at $\roots =7$~TeV},
	\IN{Phys. Rev. D}{94}{2016}{054908}
\bibitem{aliceD2}
	\BY{ALICE Collaboration}
	\TITLE{Measurement of electrons from heavy-flavour hadron decays in p--Pb collisions at $\rootsnn = 5.02$~TeV},
	\IN{Phys. Lett. B}{754}{2016}{81-93}
\bibitem{phsd}
	\BY{Song T., Cassing W., Moreau P., \atque Bratkovskaya E.}
	\TITLE{Open charm and dileptons from relativistic heavy-ion collisions},
	\IN{Phys. Rev. C}{97}{2018}{064907},
	arXiv:1803.02698 [nucl-th].
\bibitem{alice-D-pbpb-raa}
	\BY{ALICE Collaboration}
	\TITLE{Prompt D$^0$, D$^+$, and D$^{*+}$ production in Pb--Pb collisions at $\rootsnn = 5.02$~TeV},
	\IN{JHEP}{01}{2022}{174},
	arXiv:2110.09420 [nucl-ex].
\bibitem{alice-hfe-raa}
	\BY{ALICE Collaboration}
	\TITLE{Measurement of electrons from semileptonic heavy-flavour hadron decays at midrapidity in pp and Pb--Pb collisions at $\rootsnn = 5.02$~TeV},
	\IN{Phys. Lett. B}{804}{2020}{135377},
	arXiv:1910.09110 [nucl-ex].
\bibitem{na60-dca}
	\BY{NA60 Collaboration}
	\TITLE{Evidence for the production of thermal muon pairs with masses above 1 GeV/$c^2$ in 158A GeV Indium-Indium Collisions},
	\IN{Eur. Phys. J C}{59}{2009}{607-623},
\bibitem{tpc-gem}
	\BY{ALICE TPC Collaboration}
	\TITLE{The upgrade of the ALICE TPC with GEMs and continuous readout},
	\IN{JINST}{16}{2021}{P03022},
\bibitem{ITS2}
	\BY{ALICE Collaboration}
	\TITLE{Technical Design Report for the Upgrade of the ALICE Inner Tracking System},
	\IN{J. Phys. G}{41}{2014}{087001}.
\bibitem{its3}
	\BY{ALICE Collaboration}
	\TITLE{Expression of Interest for an ALICE ITS Upgrade in LS3},
	http://cds.cern.ch/record/2644611.
\bibitem{alice-pp-lowb}
	\BY{ALICE Collaboration}
	\TITLE{Soft-Dielectron Excess in Proton--Proton Collisions at $\roots = 13$~TeV},
	\IN{Phys. Rev. Lett.}{127}{2021}{042302},
	arXiv:2005.14522 [nucl-ex].
\bibitem{alice3-loi}
	\BY{ALICE Collaboration}
	\TITLE{Letter of intent for ALICE 3: A next-generation heavy-ion experiment at the LHC},
	arXiv:2211.02491 [physics.ins-det].
\bibitem{low}
	\BY{Low F.E.}
	\TITLE{Bremsstrahlung of very low-energy quanta in elementary particle collisions},
	\IN{Phys. Rev.}{110}{1958}{974-977}.
\bibitem{cond}
	\BY{Arnold P.B., Moore G.D., \atque Yaffe L.G.}
	\TITLE{Transport coefficients in high temperature gauge theories. 1. Leading log results},
	\IN{JHEP}{11}{2000}{001},

	




\end{thebibliography}
\end{document}